\documentclass{aastex}
\usepackage{spr-astr-addons}             
\usepackage{url}
\usepackage[utf8x]{inputenc}
\usepackage{lscape, morefloats, graphicx, float}   
\RequirePackage{color}                   

\begin{document}

\title{Metallicity and absolute magnitude calibrations for F-G type main-sequence stars in the {\it Gaia} era}
\slugcomment{Not to appear in Nonlearned J., 45.}
\shorttitle{Metallicity and absolute magnitude calibrations}
\shortauthors{M. {\c C}elebi, Bilir, S., Ak, S., T. Ak, Z. F. Bostanc\i, T. Yontan}

\author{M. {\c C}elebi \altaffilmark{1}}
\altaffiltext{1}{Istanbul University, Institute of Graduate 
Studies in Science, Programme of Astronomy and 
Space Sciences, 34116, Beyaz{\i}t, Istanbul, Turkey\\
Corresponding Author: murvet.celebi@gmail.com\\}

\author{S. Bilir \altaffilmark{2}} 
\altaffiltext{2}{Istanbul University, Faculty of Science, Department 
of Astronomy and Space Sciences, 34119 University, Istanbul, Turkey\\}
\and
\author{S. Ak \altaffilmark{2}}
\altaffiltext{2}{Istanbul University, Faculty of Science, Department 
of Astronomy and Space Sciences, 34119 University, Istanbul, Turkey\\}
\and
\author{T. Ak\altaffilmark{2}} 
\altaffiltext{2}{Istanbul University, Faculty of Science, Department 
of Astronomy and Space Sciences, 34119 University, Istanbul, Turkey\\}
\and
\author{Z. F. Bostanc\i\altaffilmark{2}} 
\altaffiltext{2}{Istanbul University, Faculty of Science, Department 
of Astronomy and Space Sciences, 34119 University, Istanbul, Turkey\\}

\author{T. Yontan \altaffilmark{1}}
\altaffiltext{1}{Istanbul University, Institute of Graduate 
Studies in Science, Programme of Astronomy and 
Space Sciences, 34116, Beyaz{\i}t, Istanbul, Turkey\\}

\begin{abstract} 
In this study, photometric metallicity and absolute magnitude calibrations were derived using F-G spectral type main-sequence stars in the Solar neighbourhood with precise spectroscopic, photometric and {\it Gaia} astrometric data for {\it UBV} photometry. The sample consists of 504 main-sequence stars covering the temperature, surface gravity and colour index intervals $5300<T_{eff} < 7300$ K, $\log g > 4$ (cgs) and $0.3<(B-V)_0<0.8$ mag, respectively. Stars with relative trigonometric parallax errors $\sigma_{\pi}/\pi\leq0.01$ were preferred from {\it Gaia} DR2 data for the estimation of their $M_V$ absolute magnitudes. In order to obtain calibrations, $(U-B)_0$ and $(B-V)_0$ colour indices of stars were preferred and a multi-variable second order equation was used. Calibrations are valid for main-sequence stars in the metallicity and absolute magnitude ranges $-2<{\rm [Fe/H]}<0.5$ dex and $2.5<M_V<6$ mag, respectively. The mean value and standard deviation of the differences between original and estimated values for the metal abundance and absolute magnitude are $\langle\Delta {\rm[Fe/H]}\rangle=0.00\pm0.11$ dex and $\langle\Delta M_V \rangle=0.00\pm0.22$ mag, respectively. In this work, it has been shown that more precise iron abundance and absolute magnitude values were obtained with the new calibrations, compared to previous calibrations in the literature.
\end{abstract}


\section{Introduction}
Stars with different luminosity, if they have precise  photometric, spectroscopic and astrometric data, play an important role to understand the structure, formation and evolution of the Milky Way. Specifically, F and G type main-sequence stars are used in Galactic archaeology surveys as they are long-lived objects and they contain the chemical composition of the molecule cloud in which they formed. These stars are often considered in the study of metallicity gradients and age-metallicity relations in the Galaxy disc, as well as in  testing the scenarios related to the formation of the Galaxy \citep{Sales09, Robin14, Helmi18}.

Today, data with high resolution and high signal-to-noise ratio are preferred to investigate the chemical composition of stars. Spectroscopic and astrometric data from large spectroscopic surveys such as RAVE \citep{Steinmetz06}, APOGEE \citep{Allende08}, SEGUE \citep{Yanny09}, GES \citep{Gilmore12}, LAMOST \citep{Zhao12}, GALAH \citep{deSilva15}, BRAVA \citep{Kunder12} and {\it Gaia} \citep{Gaia18} are more valuable ones. Thanks to such spectroscopic surveys, radial velocities of stars as well as atmospheric model parameters are calculated. The metallicity, which is one of these parameters, has an important role in the investigation of the chemical evolution of our Galaxy.

The metallicity of stars can be determined by spectral analysis as well as by using photometric methods. Metallicity obtained from spectral methods gives more accurate results for the stars in the Solar neighbourhood, although sensitivity in metallicity decreases as SNR becomes lower towards the fainter stars at large distances. On the other hand, fainter objects may have accurate photometric data since they can be observed at higher SNRs, i.e. Sloan Digital Sky Survey \citep[SDSS,][]{York00}. Thus, the atmospheric model parameters calculated from spectra of nearby stars can be combined with accurate photometric data to determine the metallicity of fainter stars.

The photometric metallicity estimation method has been used by many researchers since 1950s. \citet{Roman55} calculated the ultra-violet (UV) excess of about 600 F and G spectral type stars with weak metallic lines by marking them in the $U-B \times B-V$ two-colour diagram. The author found that UV excesses of the stars change in the range of 0 to 0.2 mag and the stars with the largest UV excesses have the largest space velocities. In the past, many researchers have explained the UV excess with the ``blanketing model'' \citep{Schwarzschild55, Sandage59, Wallerstein62}, in which the stars with the same metal abundance but different $B-V$ colours contain different UV excesses in the $U-B\times B-V$ two-colour diagram. \citet{Sandage69} introduced a guillotine factor to normalize UV excesses to the one for $(B-V)_0=0.6$ mag and then calculated the normalized values for different $B-V$ colours. \citet{Carney79} explained the normalized UV excesses of the main-sequence stars within 100 pc according to the iron abundance by a quadratic equation. \citet{Karaali03a} considered 77 main-sequence stars covering colour and iron abundance intervals $0.37< (B-V)_0<1.07$ mag and $-2.70<{\rm [Fe/H]}<+0.26$ dex, respectively, and calculated their UV excesses to the iron abundance by a cubic equation. \citet{Karatas06} described UV excesses of the F and G spectral type main-sequence stars in the Solar neighbourhood according to the iron abundance by a quadratic equation. \citet{Karaali11} obtained a more accurate photometric metallicity calibration using 701 main-sequence stars covering colour index and iron abundance intervals $0.32<(B-V)_0<1.16$ mag and $-1.76<{\rm [Fe/H]}<+0.40$ dex, respectively, in the {\it UBV} photometric system. Furthermore, there are several studies on photometric metallicity calibration developed for different photometric systems \citep{Walraven60, Stromgren66, Cameron85, Laird88, Buser90, Trefzger95}. The most recent photometric metallicity calibration in the {\it UBV} system is presented by \citet{Tuncel16}. In their last study, \citet{Tuncel17} obtained a photometric metallicity calibration for the SDSS photometric system using the transformation formulae of \citet{Chonis08}.

In the investigation of chemical evolution of the Milky Way, distances of stars as well as their iron abundance must be known. Distances of the stars in the vicinity of the Sun is calculated by trigonometric parallax method, while photometric parallax is preferred for the more distant stars. Using the ground-based trigonometric data of nearby stars, \citet{Laird88} calibrated differences in absolute magnitudes ($\Delta M_V$) of the field stars and Hyades stars with the same colour index, according to their UV excesses ($\delta_{0.6}$) and $(B-V)_0$ colour index of the stars. Using similar procedure and {\it Hipparcos} data \citep{ESA97}, \citet{Karaali03b} and \citet{Karatas06} developed absolute magnitude calibrations in the {\it UBV} photometric system. \citet{Karaali05}, who established the metallicity and absolute magnitude calibrations in {\it UBV} photometric system, have moved their calibrations to the SDSS system using new photometric transformation equations. \citet{Bilir05} and \citet{Juric08} adapted the photometric parallax method to the SDSS system and produced calibrations to estimate the distances of the main-sequence stars. \citet{Bilir08, Bilir09} calibrated trigonometric parallaxes of the main-sequence stars in the re-reduced {\it Hipparcos} catalogue \citep{vanLeeuwen07} to {\it UBV}, SDSS and 2MASS, and obtained absolute magnitude relations for the main-sequence stars in the range of A and M types, which are located in the Galactic disc. \citet{Tuncel16, Tuncel17} calibrated absolute magnitudes of 168 F-G dwarfs and the re-reduced {\it Hipparcos} parallaxes to {\it UBV} and SDSS photometric systems. 

Thanks to {\it Gaia} satellite data, it is aimed to create a three-dimensional map of more than one billion objects in our Galaxy and beyond, and to investigate the origin and subsequent evolution of the Milky Way. {\it Gaia} DR2 release includes astrometric data for 1.7 billion sources observed in 22 months from the beginning of the {\it Gaia} mission \citep{Gaia18}. It covers a large magnitude range with significant uncertainties in the trigonometric parallaxes. For example, astrometric uncertainties of the stars with $G$ = 15, 17, 20 and 21 mag reach 0.04, 0.1, 0.7 and 2 mas in parallax, respectively. This shows that uncertainties increase towards fainter stars. It is important to examine stars in different populations for understanding the structure and evolution of our Galaxy. Therefore, we need accurate photometric parallax relations obtained with calibration of sensitive photometric data to the {\it Gaia} astrometric data. In this study, we obtain photometric metallicity and absolute magnitude calibrations using F and G spectral type main-sequence stars in the Solar neighbourhood. In this context, this study can sort of be a supplementary work for {\it Gaia} DR2. 

Even if distances of 1.3 billion objects can be calculated from trigonometric parallax data in {\it Gaia} DR2, uncertainties in measurements reduce the sensitivity in distances. In particular, this prevents researchers investigating star fields up to faint limiting apparent magnitudes to obtain reliable Galactic model parameters or Galactic metallicity gradients. Combining precise spectroscopic, astrometric and photometric data, metal abundance and absolute magnitude calibrations to be produced for large area photometric sky surveys, i.e. SDSS, Panoramic Survey Telescope and Rapid Response System (Pan-STARRS) etc., will make a significant contribution to the more sensitive metal abundance and distance calculations of faint stars. The accuracy of the absolute magnitude calibration could be compared with the expected {\it Gaia} trigonometric parallax accuracy at the end of the mission to show that the calibration could be useful in a few years. The same for metallicity which will be provided by {\it Gaia} probably in 2021.

This work provides metallicity and absolute magnitude calibrations for investigation of not only nearby stars but also distant stars in detail. It is organized as follows: Section 2 presents the data used in the calibrations. Section 3 describes the methods for calibrations. Section 4 contains a summary and discussion.

\section{Data}

In order to obtain precise metallicity and absolute magnitude calibrations, we prioritized the stars that have sensitive spectroscopic, astrometric and photometric data in the recent studies. In this context, we used the spectroscopic data from 12 studies \citep{Boesgaard11, Nissen11, Ishigaki12, Mishenina13, Molenda13, Bensby14, daSilva15, Sitnova15, Brewer16, Maldonado16, Luck17, DelgadoMena17}. Information about the spectroscopic data used in this study is given in Table 1. We selected 5756 stars whose atmospheric model parameters ($T_{eff}$, $\log g$ and [Fe/H]) were derived in these studies. We took into account effective temperature ($T_{eff}$), surface gravity ($\log g$) and colour index $(B-V)$ given for a large spectral range of main-sequence stars. In the selection of F-G type main-sequence stars, we constrained the effective temperatures ($5300 <T_{eff}< 7300$ K) and surface gravities ($\log g \geq 4$ cgs) and composed  3413 stars in total. These limits for $T_{eff}$ and $\log g$ were taken from \citet{Eker18}. From {\it Gaia} DR2 \citep{Gaia18}, we collected trigonometric parallaxes and their errors for 3344 stars. We took into account relative parallax errors ($\sigma_{\pi}/\pi$) to obtain precise absolute magnitudes of the calibration stars. To do this, we performed the relative parallax error histogram of 3344 stars (Fig. 1). The relative parallax errors of the stars in the sample are found to extend up to $\sigma_{\pi}/\pi=0.1$. The most accurate ones are in the range $0<\sigma_{\pi}/\pi\leq0.01$ that constitute 85\% of all measurements. The number of stars decreased to 2959 by limiting the data sample with the stars whose relative parallax errors are $\sigma_{\pi}/\pi\leq0.01$. The effect of Lutz-Kelker \citep[LK,][]{Lutz73} bias on trigonometric parallaxes of the stars was also investigated. When the LK correction proposed by \citet{Smith87} was applied to programme stars with $\sigma_{\pi}/\pi\leq0.01$, an improvement of only about 0.04\% was obtained in the trigonometric parallax data. As the LK correction is too small, we did not apply the LK correction to trigonometric parallaxes in the sample.

\begin{table*}[htbp]
\setlength{\tabcolsep}{2pt}
{\tiny
  \caption{Information about spectroscopic data used in this study: Number of stars ($N$), spectral resolution ($R$), signal/noise ratio ($S/N$), and knowledge about observatory, telescope and spectrograph.}
    \begin{tabular}{clcccl}
\hline
    ID & Authors & $N$ & $R$ & $S/N$ & Observatory / Telescope / Spectrograph \\
\hline
 1 & \citet{Boesgaard11} & 117 & $\sim$42000 &   106  & Keck / Keck I / HIRES \\
 2 & \citet{Nissen11}    & 100 & 55000       & 250-500& ESO / VLT / UVES, ORM / NOT / FIES \\
 3 & \citet{Ishigaki12}  &  97 & 100000      & 140-390& NAOJ / Subaru / HDS \\
 4 & \citet{Mishenina13} & 276 & 42000       &  $>100$& Haute-Provence / 1.93m / ELODIE \\
 5 & \citet{Molenda13}   & 221 & 25000-46000 & 80-6500& ORM / NOT / FIES, OACt / 91cm / FRESCO, ORM / Mercator / HERMES\\
   &  &  &  &  & OPM / TBL / NARVAL, MKO / CFHT / ESPaDOnS \\
 6 & \citet{Bensby14}    & 714 & 40000-110000& 150-300& ESO / 1.5m and 2.2m / FEROS, ORM / NOT / SOFIN and FIES, \\
   &  &  &  &  &                                     ESO / VLT / UVES, ESO / 3.6m / HARPS, Magellan Clay / MIKE \\
 7 & \citet{daSilva15}   & 309 & $\sim$42000 & $>150$ & Haute Provence / 1.93m / ELODIE \\
 8 & \citet{Sitnova15}   &  51 & $>60000$    & 70-100 & Lick / Shane 3m / Hamilton, CFH / CFHT / ESPaDOnS \\
 9 & \citet{Brewer16}    &1617 & $\sim$70000 & $>200$ & Keck / Keck I / HIRES \\
10 & \citet{Maldonado16} & 154 & $\sim$42000-115000 & 107 & La Palma / Mercator / HERMES, ORM / NOT / FIES, \\
   &  &  &  &  & Calar Alto / 2.2m / FOCES, ORM / Nazionale Galileo / SARG \\
11 & \citet{Luck17}      &1041 & 30000-42000 & $>75$  & McDonald / 2.1m / SCES, McDonald / HET / High-Resolution \\
12 &\citet{DelgadoMena17}&1059 & $\sim$115000 & $>200$ & HARPS GTO programs \\
\hline
    \end{tabular}%
\\
HIRES: High Resolution Echelle Spectrometer, ESO: European Southern Observatory, VLT: Very Large Telescope,
UVES: Ultraviolet and Visual Echelle Spectrograph, NOT: Nordic Optical Telescope, FIES: The high-resolution Fibre-fed Echelle Spectrograph, NAOJ: National Astronomical Observatory of Japan, HDS: High Dispersion Spectrograph, ORM: Observatorio del Roque de los Muchachos, OACt: Catania Astrophysical Observatory
OPM: Observatorie Pic du Midi, FRESCO: Fiber-optic Reosc Echelle Spectrograph of Catania Observatory, TBL: Telescope Bernard Lyot, CFH: Canada-France-Hawaii, CFHT: Canada-France-Hawaii Telescope, SCES: Sandiford Cassegrain Echelle Spectrograph, HET: Hobby-Eberly Telescope, MKO: Mauna Kea Observatory, FEROS: The Fiber-fed Extended Range Optical Spectrograph, SOFIN: The Soviet-Finnish optical high-resolution spectrograph, HARPS: High Accuracy Radial velocity Planet Searcher, MIKE: Magellan Inamori Kyocera Echelle, ESPaDOnS: an Echelle SpectroPolarimetric Device for the Observation of Stars at CFHT, HERMES: High-Efficiency and high-Resolution Mercator Echelle Spectrograph, FOCES: a fibre optics Cassegrain echelle spectrograph, GTO: Guaranteed Time Observations.
  \label{tab:addlabel}%
}
\end{table*}%

\begin{figure}[t]
\begin{center}
\includegraphics[scale=0.40, angle=0]{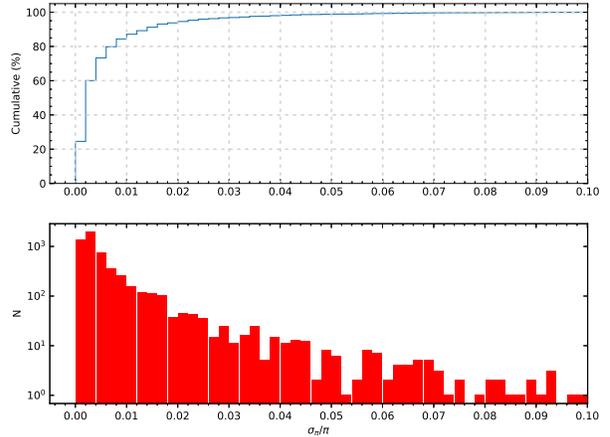}
\caption[] {The histogram of relative parallax errors, $\sigma_{\pi}/\pi$, of 3344 main-sequence stars (lower panel) and their cumulative distribution (upper panel).}
\end{center}
\end{figure}

\begin{figure}
\begin{center}
\includegraphics[scale=0.45, angle=0]{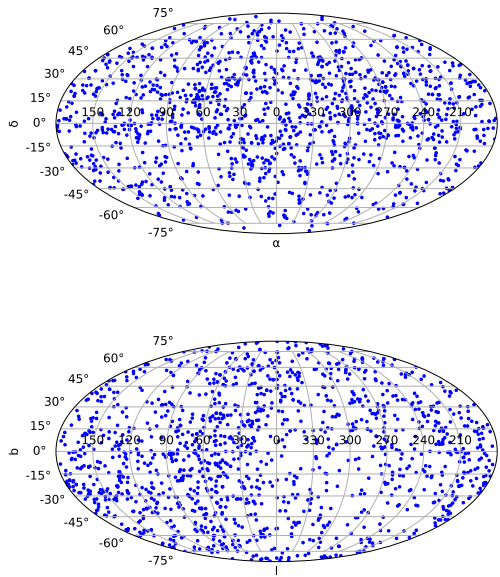}
\caption[] {The positions of 1377 F-G type main-sequence stars on the equatorial (upper panel) and Galactic (lower panel) coordinates.} 
\end{center}
\end {figure}
 
To obtain the {\it UBV} photometric data of 2959 main-sequence stars, we scanned the catalogues of \citet{Oja84, Mermilliod87, Mermilliod97, Koen10}, by considering the equatorial coordinates. Thus, we reached the {\it UBV} photometric data of 1377 stars. Distributions of stars according to equatorial and Galactic coordinates are shown in Fig. 2. The star sample was divided into three different Galactic latitude intervals, $|b|\leq30^{o}$, $30^{o}<|b|\leq 60^{o}$ and $60^{o}<|b|\leq90^{o}$. We found that the number of stars in these intervals are 693, 476, and 208, respectively. According to these results, about half of the programme stars are located at low Galactic latitudes. In this study, we used the dust map of \citet{Schlafly11} in order to remove the absorption and reddening effects of the interstellar medium. Since the absorption in a star's direction obtained from the dust map is given for the Galactic border, this absorption value should be reduced to the distance between the Sun and the relevant star. We adopted the total absorption $A_{\infty}(b)$ in {\it V}-band for a star's direction obtained from the dust map of \citet{Schlafly11}\footnote{https://irsa.ipac.caltech.edu/applications/DUST/} and estimated absorption $A_d(b)$ for the distance between Sun and the star using the following equation of \citet{Bahcall80}:

\begin{equation}
A_{d}(b)=A_{\infty}(b)\Biggl[1-\exp\Biggl(\frac{-\mid
d~\sin(b)\mid}{H}\Biggr)\Biggr].
\end{equation} 
Here, $d$ and $b$ denote the distance of the star and Galactic latitude, respectively. $H$ is the scale height of the Galactic dust \citep[$H=125$ pc;][]{Marshall06}. We calculated the distances of the stars from {\it Gaia} DR2 trigonometric parallaxes using the equation of $d=1000/\pi$ (mas). The colour excess of the star ($E_d(B-V)$) in question could be calculated from Eq. (2) \citep{Cardelli89} and the colour excess $E_d(U-B)$ was found using Eq. (3)  \citep{Garcia88}:

\begin{eqnarray}
E_d(B-V)=A_d(b)/3.1,
\end{eqnarray}
{\small
\begin{eqnarray}
E_d(U-B)=0.72\times E_d(B-V)+0.05\times E_d(B-V)^2. 
\end{eqnarray}}
The de-reddened apparent magnitude ($V_0$) and colour indices ($(U-B)_0$ and $(B-V)_0$) are then calculated by following equations,
\begin{eqnarray}
V_0= V-3.1\times E_d(B-V),\\ \nonumber
(U-B)_0=(U-B)-E_d(U-B),\\ \nonumber
(B-V)_0=(B-V)-E_d(B-V).\\ \nonumber
\end{eqnarray}
The original $E_{\infty}(B-V)$ and reduced $E_d(B-V)$ colour excesses of 1377 stars are shown in Fig. 3a and 3b, respectively.

\begin{figure}[H]
\begin{center}
\includegraphics[scale=0.40, angle=0]{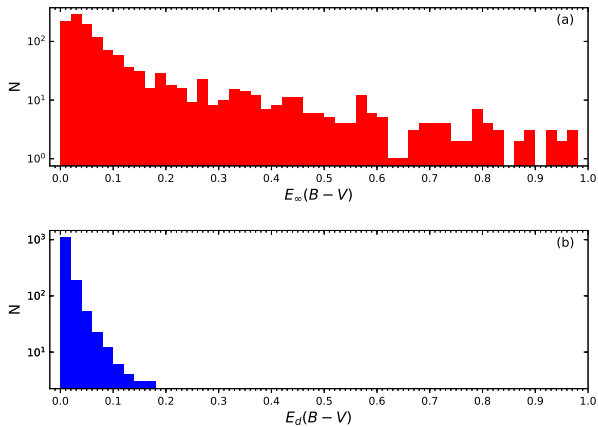}
\caption[] {Histograms of the original $E_{\infty}(B-V)$ (a) and reduced $E_d(B-V)$ (b) colour excesses of  1377 F-G type main-sequence stars.}
\end{center}
\end{figure}

Moreover, another constraint was applied to de-reddened colour excesses $(B-V)_0$ of the selected programme stars. For the photometric selection of F-G type main-sequence stars, we used the $0.3<(B-V)_0<0.8$ mag colour index interval given by \citet{Eker18}, decreasing the number of stars in the sample to 876. A final limitation to the sample has also been made according to the photometric variability of the stars. By removing the stars with registered information such as variable stars, chromospherically active stars, pulsating stars etc. in the Simbad database\footnote{http://simbad.u-strasbg.fr/simbad/}, we reduced the sample to 504 stars. The spectroscopic, photometric, and astrometric data of the 504 calibration stars are listed in Table 2.

\begin{figure}[H]
\begin{center}
\includegraphics[scale=0.52, angle=0]{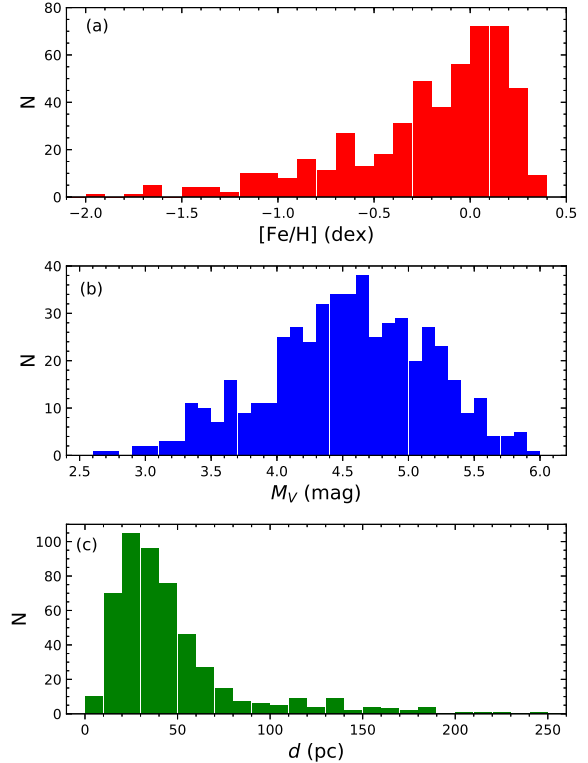}
\caption[] {The distribution of spectroscopic metal abundances (a), $M_V$ absolute magnitudes estimated from the {\it Gaia} DR2 trigonometric parallaxes (b), and distances to the Sun of the calibration stars (c).}
\end{center}
\end{figure}
\begin{table*}[htbp]
\setlength{\tabcolsep}{1.7pt}
{\scriptsize
  \centering
  \caption{The basic parameters of 504 calibration stars; ID, equatorial coordinates in J2000 ($\alpha$, $\delta$), photometric data ($V$, $U-B$, $B-V$), reduced colour excess ($E_d(B-V)$), atmospheric model parameters ($T_{eff}$, $\log g$ and [Fe/H]) and their references (Bibcode), and trigonometric parallaxes ($\pi$) taken from {\it Gaia} DR2. Parallax errors of the stars are given in brackets.}
    \begin{tabular}{clccccccccclcc}
     \hline
    ID & Star    & $\alpha$      & $\delta$     & $V$  & $U-B$ & $B-V$ & $E_d(B-V)$ & $T_{eff}$ & $\log g$      & [Fe/H] & \multicolumn{1}{c}{Bibcode} & $\pi$  & $\pi_{err}$\\
       &         & (hh:mm:ss.ss) & (dd:mm:ss.s) & (mag)& (mag) & (mag) & (mag)      & (K)       & (cm s$^{-2}$) & (dex)  &         & (mas)  & (mas)  \\
      \hline
    1  & HD 224930 & 00 02 10.34 &  +27 04 54.48 & 5.744 &  0.051 & 0.673 & 0.002 & 5454 & 4.54 &-0.73 & 2017AJ....153...21L  & 79.0696 & 0.5621\\
    2  & HD 225297 & 00 05 02.63 &$-$36 00 54.43 & 7.730 &  0.010 & 0.540 & 0.004 & 6181 & 4.55 &-0.09 & 2017A\&A...606A..94D & 19.6250 & 0.0570\\
    3  & HD 142    & 00 06 19.18 &$-$49 04 30.68 & 5.700 &  0.022 & 0.516 & 0.002 & 6403 & 4.62 & 0.09 & 2017A\&A...606A..94D & 38.1605 & 0.0648\\
    4  & HD 166    & 00 06 36.78 &  +29 01 17.41 & 6.100 &  0.325 & 0.752 & 0.002 & 5481 & 4.52 & 0.08 & 2017AJ....153...21L  & 72.5764 & 0.0498\\
    5  & HD 400    & 00 08 40.94 &  +36 37 37.65 & 6.174 & -0.066 & 0.492 & 0.005 & 6203 & 4.07 &-0.23 & 2014A\&A...562A..71B & 30.6749 & 0.0522\\
  ...  & ...       &  ...        &  ...          & ...   & ...    & ...   & ...   & ...  & ...  & ...  & ... & ... & ...\\
  ...  & ...       &  ...        &  ...          & ...   & ...    & ...   & ...   & ...  & ...  & ...  & ... & ... & ...\\
  ...  & ...       &  ...        &  ...          & ...   & ...    & ...   & ...   & ...  & ...  & ...  & ... & ... & ...\\
   499 & HD 224022 & 23 54 38.62 &$-$40 18 00.22 & 6.019 &  0.107 & 0.574 & 0.002 & 6236 & 4.46 & 0.21 & 2014A\&A...562A..71B & 35.2618 & 0.0780\\
   501 & HD 224383 & 23 57 33.52 &$-$09 38 51.07 & 7.863 &  0.146 & 0.641 & 0.011 & 5833 & 4.39 & 0.02 & 2014A\&A...562A..71B & 19.4670 & 0.0642\\
   502 & HD 224465 & 23 58 06.81 &  +50 26 51.64 & 6.640 &  0.190 & 0.665 & 0.006 & 5722 & 4.30 & 0.06 & 2015A\&A...580A..24D & 41.6957 & 0.0541\\
   503 & HD 224619 & 23 59 28.43 &$-$20 02 04.97 & 7.468 &  0.284 & 0.744 & 0.004 & 5436 & 4.39 &-0.20 & 2017A\&A...606A..94D & 38.1312 & 0.0519\\
   504 & HD 224635 & 23 59 29.29 &  +33 43 25.88 & 5.810 &  0.020 & 0.520 & 0.005 & 6227 & 4.36 & 0.03 & 2017AJ....153...21L  & 33.7083 & 0.0406\\
    \hline
        \end{tabular}%
    }
  \label{tab:addlabel}%
\end{table*}%

The histograms of the spectroscopic metallicities ([Fe/H]), the absolute magnitudes ($M_V$) estimated from {\it Gaia} DR2 trigonometric parallaxes \citep{Gaia18} using $M_V=V-5\log(1000/\pi)+A_V$ equation, and  distances ($d$) to the Sun calculated from $d=1000/\pi$ equation for 504 F-G type main-sequence stars are shown in Fig. 4. The metallicity interval of the stars is $-2<{\rm [Fe/H]}<0.5$ dex, and 81\% of the sample is concentrated in the $-0.6<{\rm [Fe/H]}<0.5$ dex metallicity interval. These results show that the vast majority of stars in the sample belong to the thin-disc population of the Galaxy \citep{Cox00}. Considering the histogram of the absolute magnitude of the sample, the stars are found in the $2.5<M_V<6$ absolute magnitude interval and frequencies of the stars are close to each other in the $4<M_V\leq5.3$ mag interval. It should be noted that the maximum distance of the sample stars to the Sun is about $d=250$ pc, while 91\% of the sample is within $d=100$ pc.

The calibration stars were divided six metal abundance intervals, and positions of the sample stars in the $(U-B)_0\times (B-V)_0$ two-colour and $M_V\times (B-V)_0$ colour-absolute magnitude diagrams were studied in accordance with metal abundances (Fig. 5). When the stars in the sample were plotted on the $(U-B)_0\times(B-V)_0$ two-colour diagram according to metallicity, it was determined that the UV excesses of the metal-poor stars are larger than the metal-rich ones (Fig. 5a). Also, in the $M_V\times(B-V)_0$ colour-absolute magnitude diagram (Fig. 5b), we found that the metal-poor stars have larger absolute magnitudes than metal-rich stars. As a result, it has been shown that metal abundance and absolute magnitude calibrations can be made by using photometric data of 504 calibration stars, which are selected according to spectroscopic, photometric and astrometric data.

\begin{figure*}
\begin{center}
\includegraphics[scale=0.6, angle=0]{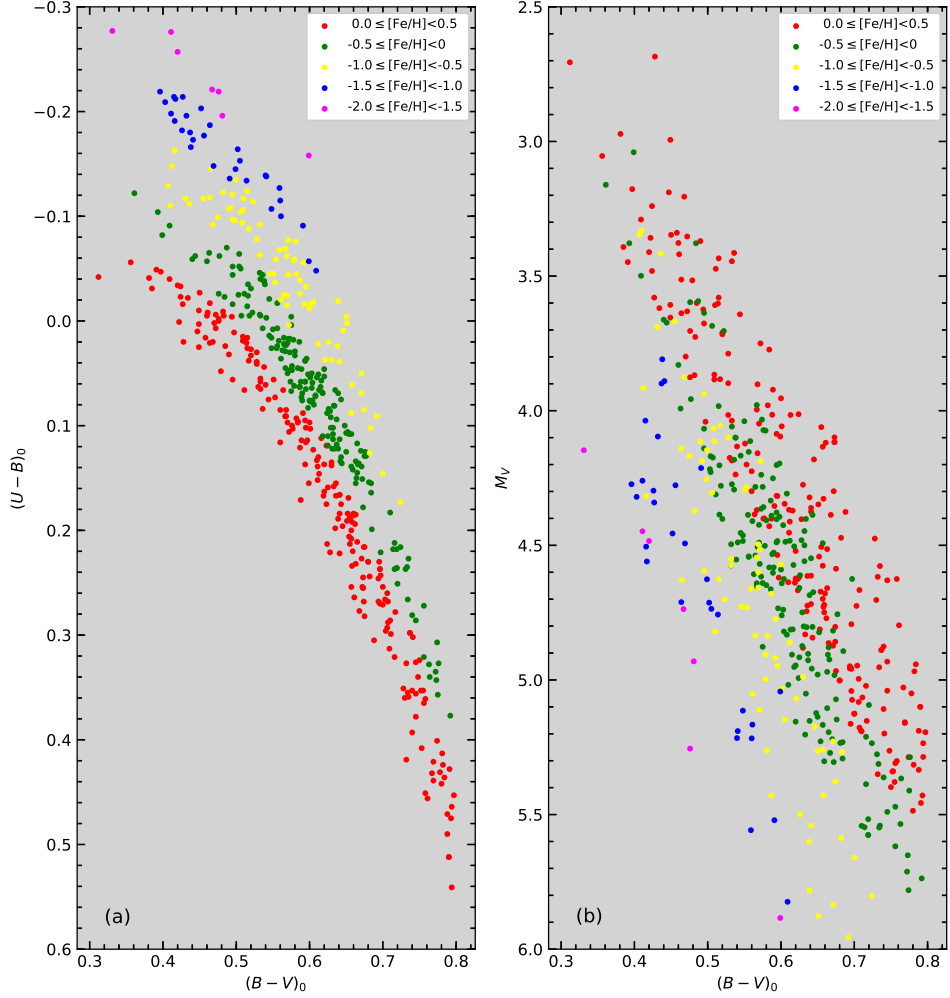}
\caption[] {The positions of the 504 main-sequence stars on the $(U-B)_0\times (B-V)_0$ two-colour (a) and $M_V\times(B-V)_0$ colour-absolute magnitude (b) diagrams according to their metal abundances.}
\end{center}
\end{figure*}

\begin{table*}
  \centering
  \caption{Coefficients and their errors for the Eq. (5). Errors  are given in brackets.}
    \begin{tabular}{ccccccc}
    \hline
    Coefficient & $a_1$ & $b_1$ & $c_1$ & $d_1$ & $e_1$ & $f_1$ \\
    \hline
         Value & -4.7503 & -8.395 & 2.141  & 5.236  & 3.775  & 0.1597  \\
               & (0.7624)& (1.476)& (1.062)& (1.118)& (1.575)& (0.0418)\\
    \hline
    \end{tabular}%
  \label{tab:addlabel}%
\end{table*}%

\section{Method}
\subsection{Metallicity calibration}

We used precise spectroscopic, photometric, and astrometric  data of 504 F-G type main-sequence stars to construct a photometric metallicity calibration. The following Eq. (5) between $(U-B)_0$ and $(B-V)_0$ colour indices and spectroscopic metal abundances of calibration stars were adopted and the parameters of variables were calculated by the multiple regression method:

{\footnotesize
\begin{eqnarray}
{\rm[Fe/H]}=a_1(U-B)_0^2+b_1(B-V)_0^2+c_1(U-B)_0(B-V)_0\\ \nonumber
+d_1(U-B)_0+e_1(B-V)_0+f_1\\\nonumber
\end{eqnarray}} 
The calculated parameters and their errors are listed in Table 3. The correlation coefficient and standard deviation of the metallicity calibration (Eq. 5) are $R^2=0.932$ and $\sigma= 0.11$ dex, respectively. Comparison of estimated photometric metal abundances (${\rm [Fe/H]}_{est}$) with original spectroscopic metal abundances (${\rm [Fe/H]}_{org}$) and differences between these values ($\Delta{\rm [Fe/H]}={\rm [Fe/H]}_{org}-{\rm [Fe/H]}_{est}$) are shown in Fig. 6. The mean value and standard deviation of the differences between estimated and original metal abundances are $\langle{\rm[Fe/H]}\rangle=0.00$ and $\sigma_{\rm[Fe/H]}=0.11$ dex, respectively. The upper panel of Fig. 6 shows that the distribution becomes denser around the one-to-one line and the scattering of the metal abundance differences is relatively small. The lower panel of Fig. 6 represents that metal abundance differences between estimated from the above calibration equation (Eq. 5) and spectroscopic data are in the range of $\pm0.26$ dex and most of the difference values are within $\pm1\sigma$.

\begin{table*}
  \centering
  \caption{Coefficients and their errors for the Eq. (6). Errors are given in brackets.}
    \begin{tabular}{ccccccc}
    \hline
    Coefficient & $a_2$ & $b_2$ & $c_2$ & $d_2$ & $e_2$ & $f_2$ \\
    \hline
    Value &  4.382  &  17.242 & -10.620 & 0.547   & -7.885  & 3.4576   \\
          & (1.477) & (2.861) & (3.996) & (0.167) & (3.052) & (0.8104) \\
    \hline
    \end{tabular}%
  \label{tab:addlabel}%
\end{table*}%

\begin{figure}[h]
\begin{center}
\includegraphics[scale=0.45, angle=0]{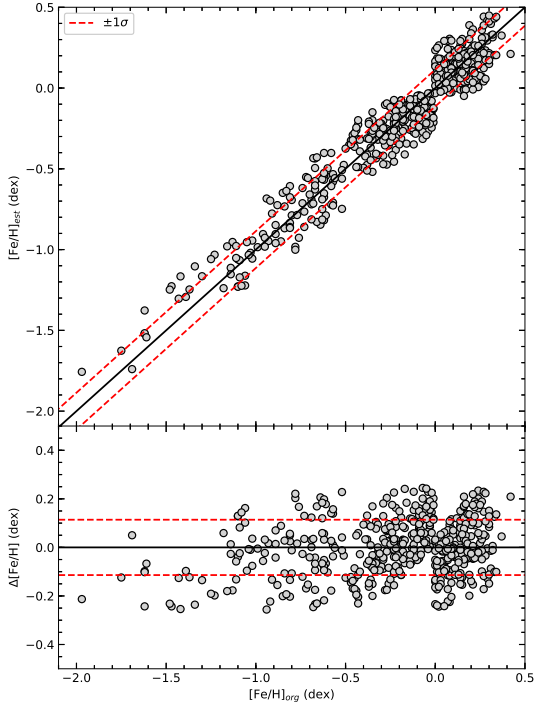}
\caption[] {Comparison of the estimated (${\rm[Fe/H]}_{est}$, photometric) and the original (${\rm[Fe/H]}_{org}$, spectroscopic) metallicities (upper panel) and distribution of the metallicity residuals $\Delta$[Fe/H] with respect to the original metallicities (lower panel) for 504 calibration stars. The black solid line represents one-to-one line and dashed red lines denote $\pm1\sigma$ prediction levels.}
\end{center}
\end{figure}

\begin{figure}[h]
\begin{center}
\includegraphics[scale=0.45, angle=0]{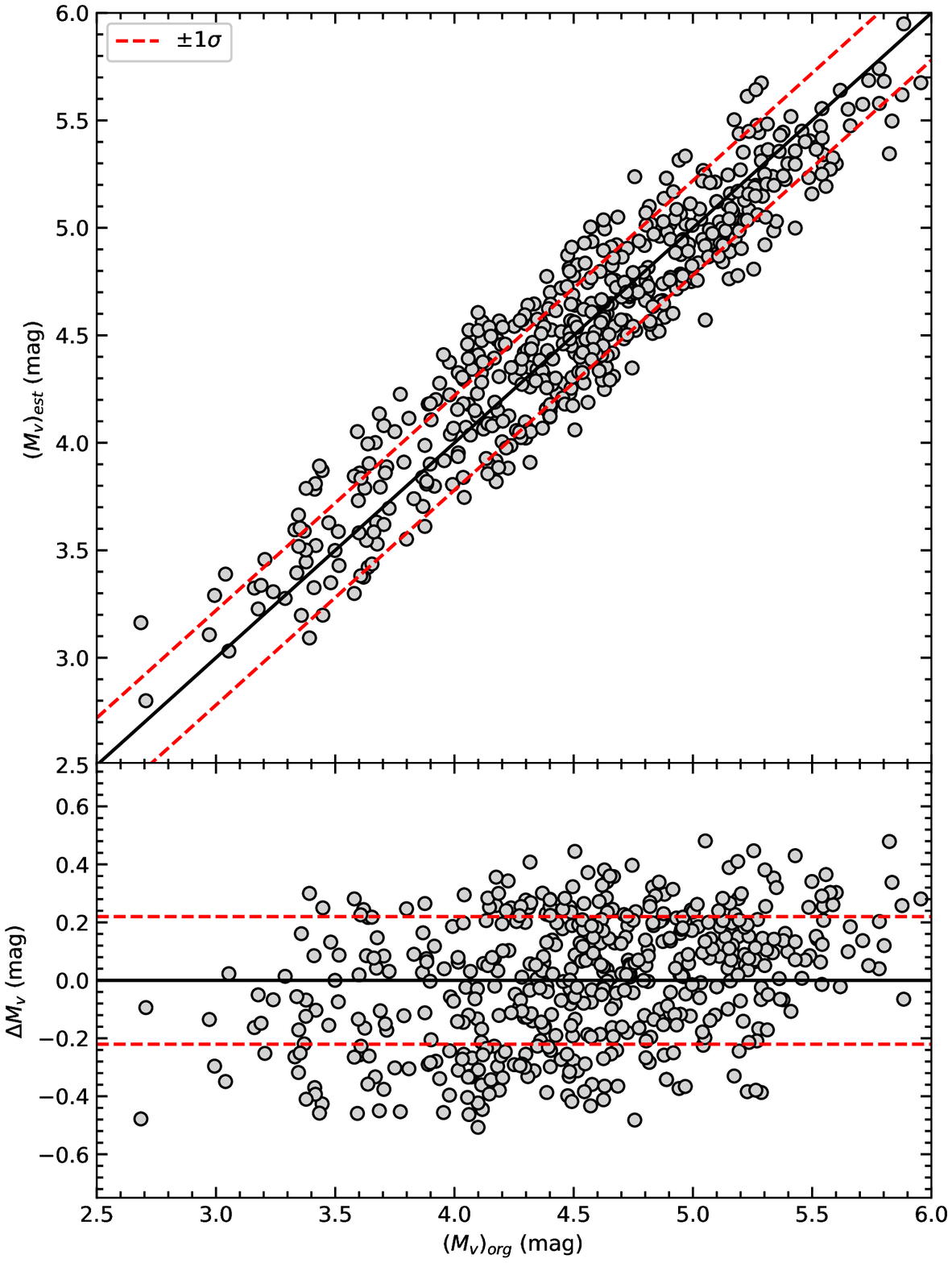}
\caption[] {Comparison of the estimated ($(M_V)_{est}$ photometric) and the original ($(M_V)_{org}$, {\it Gaia} trigonometric parallaxes) absolute magnitudes (upper panel) and distribution of the absolute magnitude residuals $\Delta M_V$  with respect to the original absolute magnitudes (lower panel) for 504 calibration stars. The black solid line represents one-to-one line and dashed red lines denote $\pm1\sigma$ prediction levels.}
\end{center}
\end{figure}

\subsection{Absolute magnitude calibration}
To construct an absolute magnitude ($M_V$) calibration in the {\it UBV} photometric system, we used 504 F-G type main-sequence stars in our data sample. The absolute magnitudes of these calibration stars were obtained with the equation, $M_V=V-5\log(1000/\pi)+A_V$, and the distances of stars in this equation were calculated from the {\it Gaia} DR2 trigonometric parallaxes \citep{Gaia18}. The following Eq. (6) between $(U-B)_0$ and $(B-V)_0$ colour indices and absolute magnitude of calibration stars was adopted and the parameters of variables were calculated by the multiple regression method:    

{\footnotesize
\begin{eqnarray}
M_V=a_2(U-B)_0^2+b_2(B-V)_0^2+c_2(U-B)_0(B-V)_0\\ \nonumber
+d_2(U-B)_0+e_2(B-V)_0+f_2\\\nonumber
\end{eqnarray}} 
The calculated parameters and their errors are listed in Table 3. The correlation coefficient and standard deviation of the absolute magnitude calibration (Eq. 6) are $R^2=0.869$ and $\sigma=0.22$ mag, respectively. Comparison of estimated absolute magnitude ($(M_V)_{est}$) with original (from {\it Gaia} trigonometric parallaxes) absolute magnitudes ($(M_V)_{org}$) and differences between these values ($\Delta M_V=(M_V)_{org}-(M_V)_{est}$) are shown in Fig. 7. The mean value and standard deviation of the differences between estimated and original absolute magnitudes are $\langle M_V \rangle=0.00$ and $\sigma_{\langle M_V \rangle}=0.22$ mag, respectively. The upper panel of Fig. 7 shows that the distribution becomes denser around the one-to-one line and the scattering of the absolute magnitude differences is relatively small. The lower panel of Fig. 7 represents that absolute magnitude differences calculated from two different methods are in the range of $\pm0.51$ mag and most of the differences are within $\pm1\sigma$.

\begin{figure*}[t]
\begin{center}
\includegraphics[scale=0.6, angle=0]{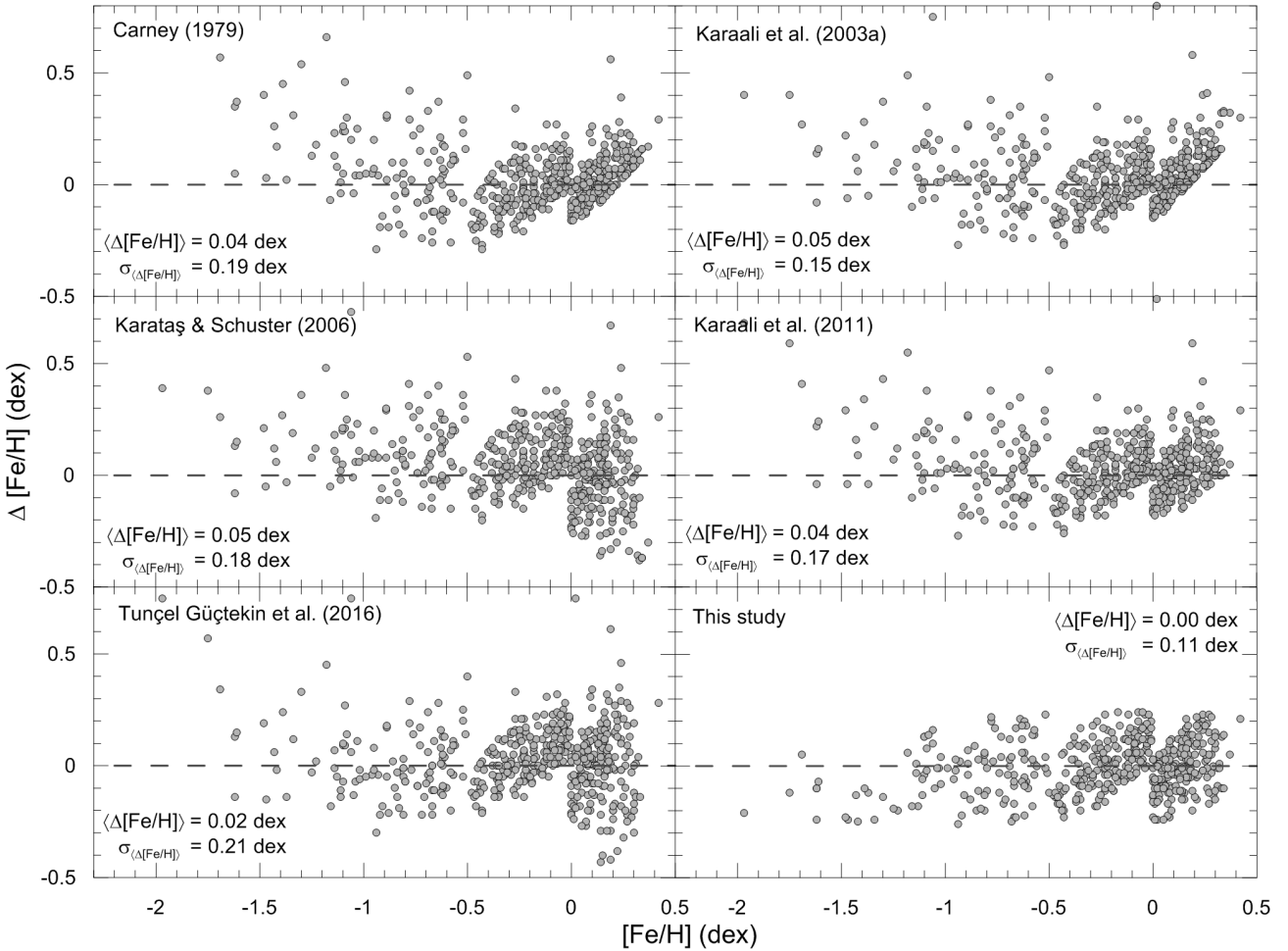}
\caption[] {Distributions of the metallicity residuals of various calibration equations estimated for 504 F-G spectral type main-sequence stars with respect to the metallicities in this study. The reference for the calibration in question is indicated in each panel, together with mean value and standard deviation of residuals.}
\end{center}
\end{figure*}

\section{Summary and Discussion}

Using precise spectroscopic ($T_{eff}$, $\log g$ and [Fe/H]), photometric ($V$, $U-B$, $B-V$) and astrometric ($\pi$) data of F and G spectral type main-sequence stars, new calibrations for the {\it UBV} photometry were obtained to estimate the metallicities and absolute magnitudes photometrically. Tough limitations on spectroscopic, photometric, and astrometric data were made to select the calibration stars, reducing the number of stars in the sample from 5756 to 504. Multiple regression method was used to construct new metal abundance and absolute magnitude calibrations, which would be a connection between the photometric data of the stars, metal abundances from spectroscopic studies and trigonometric parallaxes determined by astrometric methods. When calculating metal abundances and absolute magnitudes of F-G type main sequence stars in low Galactic latitudes using photometric calibrations, it should be noted that they are exposed to extreme interstellar medium and the colour excess should be sensitively determined.

\subsection{Comparison of new metallicity calibration with those in the literature}
In the literature, the relationships with polynomials of different degrees between the UV residuals of the stars and spectroscopic metallicities were used. For about 60 years, the application of similar methods to main-sequence stars, increasing photometric, spectroscopic and astrometric sensitivities with the developing technology have affected sensitivities of the calibrations. In the literature, five studies attract our attention when the UV residuals-sensitive metallicity calibrations for the {\it UBV} photometry are investigated: \citet{Carney79}, \citet{Karaali03a}, \citet{Karatas06}, \citet{Karaali11}, and \citet{Tuncel16}. Analytical expressions of the metallicity calibrations in these studies reveal that the colour index of the calibrations depends on the UV residuals ($\delta_{0.6}$) normalized to $(B-V)_0=0.6$ mag. The primary difference between the metallicity calibrations is that the photometric UV residuals of the stars are expressed in different degrees of polynomials. We applied the five aforementioned metallicity calibrations to 504 stars in our sample for a further comparison of our results with the ones in these studies. The distributions of differences ($\Delta{\rm [Fe/H]}$) between photometric metallicities estimated from the calibration in each study and the original spectroscopic metallicities are shown in Fig. 8 according to the  spectroscopic ones (${\rm [Fe/H]}_{org}$).

Note that similar distributions are obtained by taking into account the validity limits of the metallicity calibrations. While the means of metallicity differences ($\Delta{\rm [Fe/H]}$) in four studies \citep{Carney79, Karaali03a, Karaali11, Karatas06} are between 0.04 and 0.05 dex, these values are lower than 0.02 dex for two studies \citep[and this study]{Tuncel16}. As for the dispersion of metallicity distributions, the smallest standard deviation value is obtained in this study, $\sigma=0.11$ dex. The standard deviations of metallicity differences range from 0.15 to 0.21 dex in previous studies. Thus, we conclude that the metallicity calibration derived in this study is more reliable than those in the literature, considering the mean metallicity differences and standard deviations. 

\begin{figure}[h]
\begin{center}
\includegraphics[scale=0.45, angle=0]{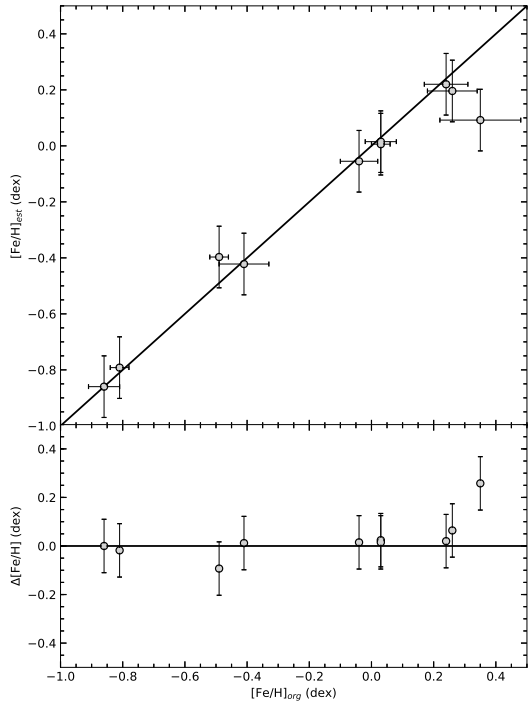}
\caption[] {Comparison of the estimated (${\rm[Fe/H]}_{est}$) and the original (${\rm[Fe/H]}_{org}$) metallicities of 10 {\it Gaia} benchmark stars which are taken from \citet{Jofre15}.}
\end{center}
\end{figure}

Another study to compare our photometric metallicity calibrations was also described by \citet{Jofre15}. They obtained iron and alpha element abundances of 34 {\it Gaia} FGK benchmark stars using eight different spectral analysis methods. Since the spectral resolution and $S/N$ of the 34 stars used in their study were quite good, the element abundances of the stars were determined with high sensitivity. In order to see the sensitivity of the new metallicity calibration produced in our study, these {\it Gaia} benchmark stars are taken into consideration. In this comparison, we used 10 main-sequence stars within the range of $-2<{\rm[Fe/H]}<0.5$ dex metallicity interval and {\it UBV} photometric data available. Photometric metal abundances were determined for these stars using our methods described in Sections 2 and 3. Comparison of estimated photometric metal abundance with spectroscopic ones are shown in Fig. 9. As shown in this figure, metal abundances of 10 stars obtained from two different methods are quite compatible with each other within the errors, except for $\mu$ Ara which has the richest metallicity in the sample. Considering 47 spectroscopic studies of $\mu$ Ara in the literature, it is seen that the metal abundance is in the range of $0.16<[{\rm Fe/H}]<0.41$ dex \citep{Francois86, Hearnshaw75} and the median metallicity of the star is [Fe/H]=0.28 dex. The median metallicity of the star and the photometric metal abundance calculated in this study ([Fe/H]=0.10 dex) is more compatible within the errors. This result shows that our photometric metallicity calibration can be used in the calculation of metal abundances of F-G main-sequence stars with an accuracy of $\pm0.11$ dex in the range $-2<{\rm[Fe/H]}<0.5$ dex, where it is valid.

Our new metallicity calibration has two superior aspects compared to similar studies in the literature. At first, it uses a large number of main-sequence stars having precise spectroscopic, photometric and astrometric data, and the other one is that it is possible to determine the photometric metallicity of F-G spectral type main-sequence stars without depending on a standardized UV residual. 

\begin{figure*}[h]
\begin{center}
\includegraphics[scale=0.54, angle=0]{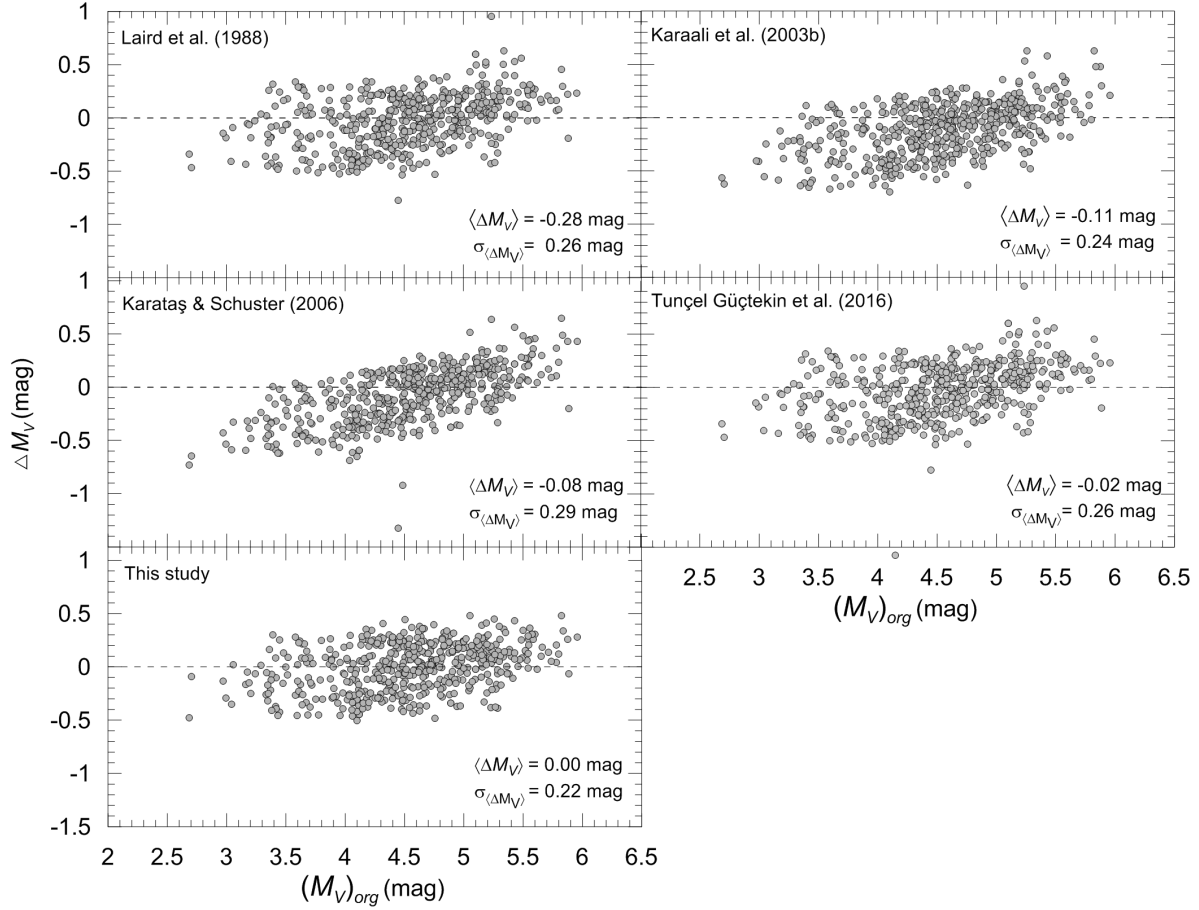}
\caption[] {Distributions of the absolute magnitude residuals of various calibration equations estimated for 504 F-G spectral type main-sequence stars with respect to the absolute magnitudes in this study. The reference for the calibration in question is indicated in each panel, together with mean value and standard deviation of residuals.}
\end{center}
\end{figure*}

\subsection{Comparison of new absolute magnitude calibration with those in the literature}

In the literature, there are four studies regarding to the UV residuals-sensitive absolute magnitude calibrations: \citet{Laird88}, \citet{Karaali03b}, \citet{Karatas06} and \citet{Tuncel16}. Common aspects of these studies, in which different photometric and trigonometric parallax data obtained in different times were used, are that their calibrations are expressed by third degree polynomial functions, they depend on UV residuals and that the main-sequence members of the Hyades open cluster are used as reference in the calculations.

The absolute magnitude calibration in this study was compared with the calibrations of the four research groups in the literature. For comparisons, 504 F-G spectral type main-sequence stars in our study were placed in the calibration relation of each study and the absolute magnitudes of the stars were calculated. The differences $\Delta M_V$ between the absolute magnitudes calculated from {\it Gaia} parallaxes and those obtained from the calibrations are shown in Fig. 10. The absolute magnitudes of the stars calculated from the calibrations are in the $2.5<M_V<6$ mag interval and the differences are found to be in the $-1.5< \Delta M_V<+1$ mag interval. In general, the calculated absolute magnitude differences of the stars in $M_V<3$ and $M_V>5.5$ mag contain systematic errors in the previous calibrations. The absolute magnitude differences calculated from the calibration of \citet{Karatas06} are compatible with the {\it Gaia} DR2 only for a very limited range of absolute magnitudes, containing larger systematic errors. The distribution of absolute magnitude differences reveal that the most incompatible calibration study was performed by \citet{Laird88}, while the calibrations of \citet{Karaali03b} and \citet{Karatas06} seem to follow it. The results of this study and \citet{Tuncel16} presents the most compatible absolute magnitudes with those of calculated from {\it Gaia} DR2, compared to the absolute magnitudes estimated from the previous calibrations. The mean values and standard deviations of the absolute magnitude differences are -0.28 and 0.26 mag, -0.11 and 0.24 mag, -0.08 and 0.29 mag, -0.02 and 0.26 mag, and 0.02 and 0.22 mag, for \citet{Laird88}, \citet{Karaali03b}, \citet{Karatas06}, \citet{Tuncel16} and this study, respectively. In terms of the mean values and standard deviations of absolute magnitude differences, we conclude that the calibration suggested in this study is very reliable. 

The application of new calibrations to the main-sequence stars can be used to predict metal abundances and distances, and in the investigation of the formation and evolution of our Galaxy \citep{Karaali03a, Karaali03b, Bilir05, Ak07, Juric08, Ivezic08, Siegel09, Karaali11, Tuncel16, Tuncel17, Tuncel19}. In addition, it could be preferable to determine the mean metallicities and distances of clusters by determining metallicities of member stars of open and globular clusters \citep[see][]{Karaali03c, Karaali14, Bilir06, Bilir10, Bilir13, Bilir16, Yontan15, Yontan19, Bostanci15, Bostanci18, Ak16}.

\section{Acknowledgments}
We thank the anonymous referees for their insightful and constructive suggestions, which significantly improved the paper. This research has made use of the SIMBAD, NASA\rq s Astrophysics Data System Bibliographic Services and the NASA/IPAC Infrared Science Archive, which is operated by the Jet Propulsion Laboratory, California Institute of Technology, under contract with the National Aeronautics and Space Administration. This work has made use of data from  the European Space Agency (ESA) mission {\it Gaia} (\mbox{https://www.cosmos.esa.int/gaia}), processed by the {\it Gaia} Data Processing and Analysis Consortium (DPAC\footnote{https://www.cosmos.esa.int/web/gaia/dpac/consortium}). Funding for the DPAC has been provided by national institutions, in particular the institutions participating in the {\it Gaia} Multilateral Agreement.

\end{document}